\documentclass[twocolumn]{emulateapj}

\usepackage{amsmath}
\usepackage{epsf}
\usepackage{psfig}
\usepackage{graphicx}
\usepackage{natbib}

\newcommand{\bc}{\begin{center}}
\newcommand{\ec}{\end{center}}

\newcommand{\kms}{km~s$^{-1}$ }
\newcommand{\hii}{HII }
\newcommand{\pa}{P$\alpha$ }

\begin{document}

\slugcomment{}

\title{Properties of the Compact \hii Region Complex G-0.02-0.07}

\author{E. Mills, M. R.  Morris}
\affil{Department of Physics and Astronomy, University of California, Los Angeles}
\email{millsb@astro.ucla.edu}

\author{C. C. Lang}
\affil{Physics and Astronomy Department, University of Iowa}

\author{H.Dong, Q. D. Wang}
\affil{Astronomy Department, University of Massachusetts, Amherst}

\author{A. Cotera}
\affil{SETI Institute, 515 N. Whisman Road - Mountain View, CA 94043}

\author {S. R. Stolovy}
\affil{Spitzer Science Center, MS 220-6, Pasadena, CA 91125}

\begin{abstract}

We present new extinction maps and high-resolution Paschen-alpha images of 
G-0.02-0.07, a complex of compact \hii regions located adjacent to the M-0.02-0.07 giant molecular cloud, 
6 parsecs in projection from the center of the Galaxy. These \hii regions, which lie in projection just outside the boundary of the Sgr A East supernova remnant, represent one of the most recent episodes of star formation in the central parsecs of the Galaxy.  The 1.87 $\mu$m extinctions of regions A, B and C are almost identical,  
 approximately 1.5 magnitudes. Region D, in contrast, has a peak extinction of  A$_{1.87}$ = 2.3 magnitudes.  Adopting the Nishiyama et al. (2008) extinction law, we find these extinctions correspond to visual extinctions of A$_{V}$ = 44.5  and A$_{V}$ = 70. The similar and uniform extinctions of regions A, B and C are consistent with that expected for foreground extinction in the direction of the Galactic center, suggesting that they lie at the front side of the M-0.02-0.07 molecular cloud. Region D is more compact, has a higher extinction and is thus suspected to be younger and embedded in a dense core in a compressed ridge on the western edge of this cloud.

\end{abstract}


\section{Introduction}

%Catchy intro

\begin{figure*}
\epsscale{1.0}
\plotone{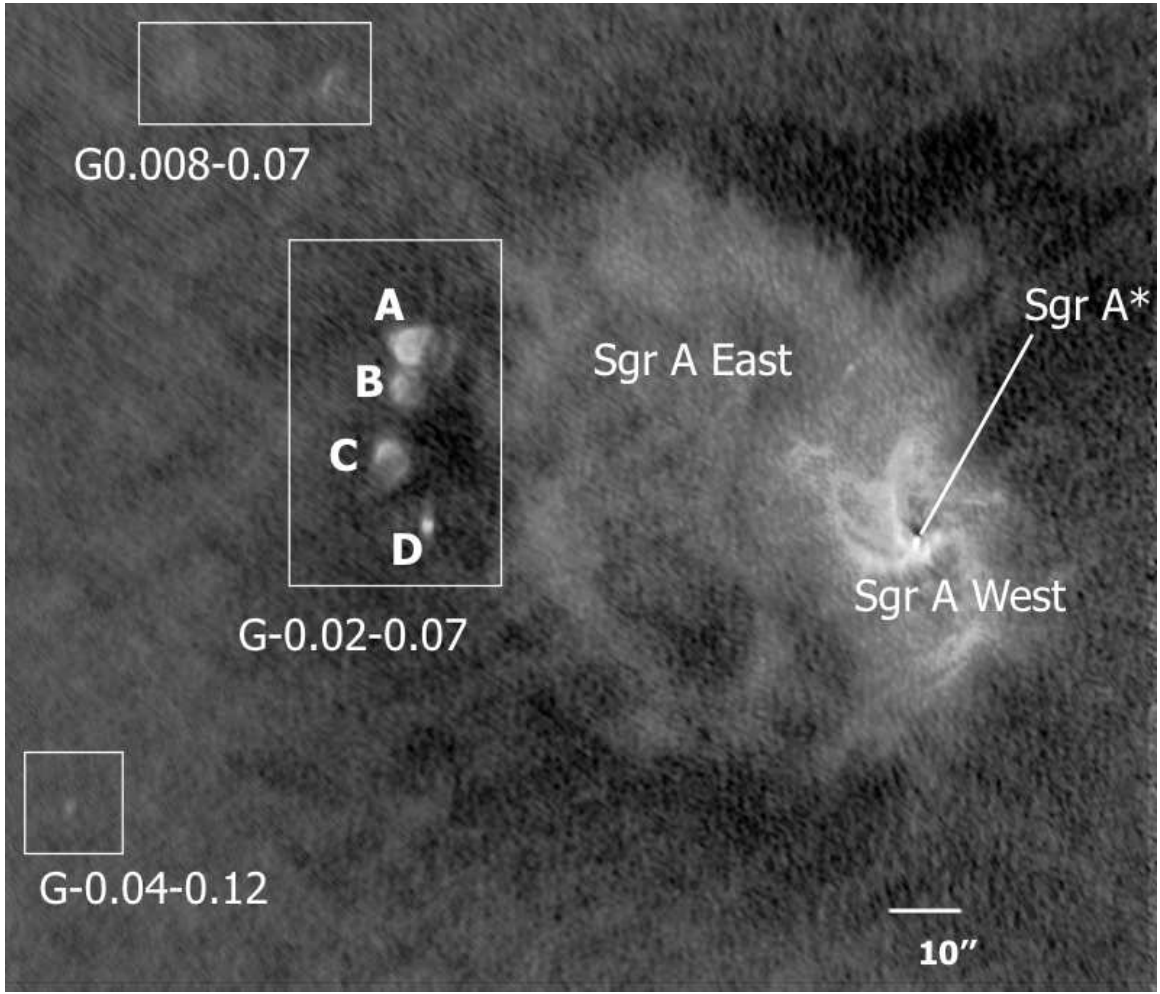}
\caption{ 8.4 GHz image showing the G-0.02-0.07 \hii regions (A,B,C and D), the Sgr A East supernova remnant, and the Sgr A West complex of ionized gas surrounding the bright point source Sgr A*, the Galaxy's central supermassive black hole. Also shown are two previously unknown radio sources, G0.008-0.07 and G-0.04-0.12.  In this image, North is up, and East is to the Left.}
\label{radiofig}
\end{figure*}

G-0.02-0.07 is a group of 4 \hii regions (three compact and one ultracompact) in the Galactic center (hereafter GC) which lie $\sim6$ parsecs in projection from the central supermassive black hole. Individually, the regions are also identified as Sgr A-A through Sgr A-D, as they were first identified in radio images of the Sgr A complex of ionized gas surrounding Sgr A*, the radio counterpart of the central black hole \citep{Ekers83}. All four \hii regions are projected to lie along the edge of the Sgr A East supernova remnant (see Figure \ref{radiofig}) which lies between these regions and Sgr A*. 

Several studies have been made of the G-0.02-0.07 complex, both in radio continuum and recombination lines \citep{Ekers83, Goss85}, as well as in mid infrared fine structure lines  \citep[hereafter S92]{Serabyn92}, including the recent work of \cite{YZ10}.  These observations have shown that the regions are consistent with each being ionized by a single late O-type star. The radial velocities of all four \hii regions have also been measured to be very similar, ranging from 43 to 49 \kms, indicating that these regions are kinematically associated both with each other, and with M-0.02-0.07, the nearby 50 km s$^{-1}$ cloud. The \hii regions appear to lie along a spatially coincident dense ridge of the M-0.02-0.07 cloud, denoted the \lq\lq molecular ridge\rq\rq by \cite{Coil00}, which shows evidence of interaction with the Sgr A East supernova remnant \citep{Serabyn92, YZ96, SP08}. Despite the suggestive arrangement of the \hii regions along the periphery of Sgr A East, estimates of its age \citep[$10^3-10^4$ years:][]{Fryer06, Mezger89} suggest that the star formation event that produced the G-0.02-0.07 complex predates the explosion, as the lifetimes of ultracompact \hii regions, precursors to compact \hii regions, are believed to span $10^5$ years \citep{WC89}.  In particular, region D, likely the youngest of the four, has a minimum age of at least a few times $10^4$ years, estimated from the mass loss rate of the central star, and the expansion rate of the nebula \citep{YZ10}.

Although these regions are thus unlikely to be an example of supernova-triggered star formation, they are valuable to study not only as the closest episode of recent (within $0.1-1$ Myrs) massive star formation to the center of the Galaxy, but also as one of very few examples of recent massive star formation in the central hundred parsecs. The central hundred parsecs of the galaxy are estimated to have a star formation rate of at least $0.05$ M$_{\odot}$yr$^{-1}$ \citep{Gusten89}, and likely higher \citep{YZ09,Schuller05}, are host to three young star clusters with initial masses in excess of $10^4$ M$_{\odot}$ \citep{Figer99, Figer02, Schod09}, and at least $4\times10^6$ solar masses of molecular material \citep{LZM02}. However the G-0.02-0.07 complex of \hii regions is one of the few regions \citep[along with a single compact \hii region in the nearby cloud M-0.13-0.08 and a complex of \hii regions in the -30 km/s cloud; ][]{Ho85, Zhao93}, of apparent massive star formation associated with the massive but largely quiescent giant molecular clouds interior (R$\,< 120$ pc) to the active star formation regions Sgr B2 and Sgr C.

We have used a combination of new infrared data and archival radio data to study these \hii regions in greater detail, and to better determine their location and relationship with the M-0.02-0.07 cloud and Sgr A East remnant. In this paper we present  high resolution (0\arcsec.2) images of this complex obtained with HST-NICMOS in the 1.87 $\mu$m Paschen $\alpha$ (hereafter \pa) line, showing the fine filamentary structures and unusual morphologies of these regions in new detail. We  also present the first maps of the extinction structure within the G-0.02-0.07 \hii regions, made from a comparison of the \pa and 8.4 GHz radio data, at arcsecond resolution. Finally, we compare our extinction measurements and morphological study of these regions with recent measurements of their gas dynamics, and discuss unusual features of two of the \hii regions, regions A and D, in more detail. 

\section{Observations and Data Reduction}

Two main data sets were analyzed for this paper: an emission line map of the near-infrared \pa (n = 4--3) recombination transition of hydrogen, and 8.4 GHz data obtained from the archives of the Very Large Array radio interferometer in New Mexico.  

\subsection{NICMOS Paschen $\alpha$ Observations}
The \pa emission line map of the G-0.02-0.07 region (Figure \ref{palpha}) is part of a larger survey of the inner 39\arcmin\, by 15\arcmin\, of the Galaxy in this line \citep{Wang09} using data from 144 orbits of the Hubble Space Telescope between February and June 2008. Observations were made with the NIC3 camera in both the F187N and F190N narrowband (1$\%$ bandpass) filters, one of which is centered on the \pa line at 1.87 $\mu$m, and the other centered on the nearby continuum at 1.90 $\mu$m. The native pixel size of NIC3 at these wavelengths is $\sim 0.2$ arcsec$^2$, which leads to an undersampled PSF. The images were dithered to achieve a sub-pixel resolution of 0\arcsec.2 (0\arcsec.1 pixels). The resulting sensitivity of these images is 0.13 mJy arcsec$^{-2}$. The data reduction and the mosaicking process used to make the map of the survey region is described in more detail in a separate paper (Dong et al. 2011).

\subsubsection{Continuum Subtraction}

In order to produce a map of pure \pa line emission, it is necessary to remove the primarily stellar continuum emission also observed in the filter. In principle, this is accomplished with duplicate observations in a neighboring narrowband filter sampling only the continuum (in this case, separated by 3000 km/s, which is sufficient to ensure there is no Doppler shifting of line emission into the continuum filter), and differencing the two images. However, the ratio of continuum emission in the two filters depends on both the spectral type of the star, and to a greater extent, the local extinction, which is highly variable toward the GC. In comparison to the reddening from the high extinction toward the GC, the former effect is negligible here. To take the effects of extinction into account, an adaptive F187N/F190N ratio is calculated over the map: the colors of the nearest 101 stars are averaged in 0\arcsec.4 by 0\arcsec.4  boxes. The majority of these stars are assumed to lie at the distance of the GC. However, where the stars are too few, due to attenuation by dense molecular clouds, and cannot be generally assumed to be located at the distance of the GC, the extinction map of \cite{Schul09} is used to determine the ratio. Additional details of this process are discussed in Dong et al., (2011).

\subsubsection{Flux Calibration}

After we flat-field the data and remove the instrumental background, we transform them to an absolute flux scale by applying a standard conversion factor from ADU s$^{-1}$ to Jy, obtained via observations of two principle calibration stars, which is assumed valid for all NIC3 data. Since the \pa images are pure line emission, their natural unit is line flux per pixel, or erg cm$^{-2}$ s$^{-1}$ pixel$^{-1}$. To convert the \pa images to units of line flux per pixel, we then multiply the flux density per pixel in Jy by the width of the F190N filter, as in \cite{Scoville03}.

\subsection{Archival VLA Data}

We obtained radio continuum data of the G-0.02-0.07 \hii regions from the archive of the Very Large Array (VLA) Radio Telescope of the National Radio Astronomy Observatory \footnotemark[1]\footnotetext[1]{The National Radio Astronomy Observatory is a facility of the National Science Foundation operated under cooperative agreement by Associated Universities, Inc.}.  
The observations were made during 1991-1992 in three array configurations (D, C and A/B (antenna move time)) and 
the integration time on source was $\sim$1 hour in each array configuration. The field of view of these observations is centered on RA, DEC (J2000): (17$^{\textrm{h}}$ 45$^{\textrm{m}}$ 51.7$^{\textrm{s}}$, -28 59\arcmin\, 23.7\arcsec), which is $\sim$6\arcsec~to the NE of region A. These data were taken in standard continuum mode (four IFs, 50 MHz per IF).

\subsubsection{Calibration and Imaging}

The data from the D,C and A/B configurations were calibrated using the standard AIPS software packages of the NRAO. The data were combined and images using IMAGR, and self-calibrated using Sgr A* as a reference source.  The resulting final image has an RMS noise of 0.2 mJy/beam, and a dynamic range of 1300. The angular resolution of this image was 1.5\arcsec x 0\arcsec.8, PA = -4.78 degrees. The overall noise was $\sim$10 times the theoretical expectation for a point source,  which is reasonable, given the complicated source structure, and the presence of the bright ($\sim$20 Jy), extended structure of Sgr A East and West at the edge of the beam. The final image does, however, display a low-level pattern of linear artifacts, or striping, running from southwest to northeast, the origin of which could not be identified in the UV data. This structure is faint enough not to affect our analysis. 

The D-array data contribute many short spacings to the (u,v) coverage; the shortest spacing in the (u,v) data used to make the final images is 0.75 kilolambda.  At 8.4 GHz, this should lead to sensitivity to structures with sizes less than 5\arcmin.6. The largest (north-south) extent of G-0.02-0.07 is only 1\arcmin, and thus we are satisfied that there should be no flux missing from our measurements of these regions beyond the smooth synchrotron background of the GC.

\section{\hii Region Properties from Radio and \pa}

The \pa images offer an unprecedented look at the detailed morphologies of this group of compact \hii regions. At an assumed distance of approximately that of Sgr A*, \citep[8.4 kpc;][]{Reid09b,Ghez08}, the angular resolution of 0\arcsec.2 corresponds to a spatial resolution of 0.008 pc ($\sim$1600 AU), higher resolution than any existing radio studies of these \hii regions. The resolution of the \pa images allows us to identify new structures including knots, filaments, diffuse ridges, and the detailed shape of the boundaries of these \hii regions. An image of the entire complex is shown in Figure \ref{palpha}, with a logarithmic stretch in order to emphasize the diffuse structure of these regions.

\subsection{Regions A,B, and C}
\label{ABC}
\begin{figure*}
\epsscale{0.85}
\plotone{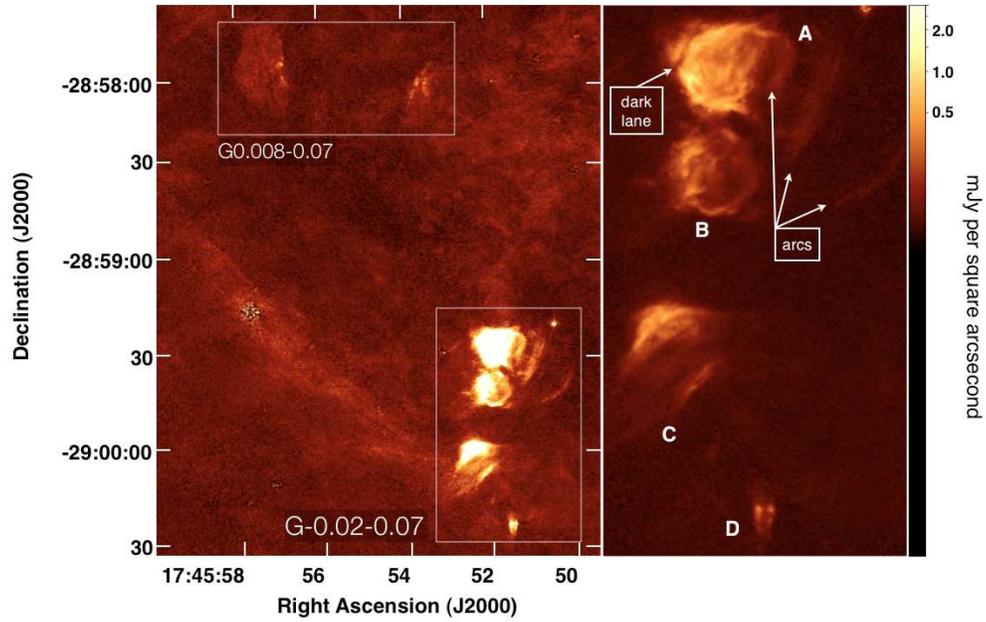}
\caption{ Left: \pa image showing a view of the G-0.02-0.07 \hii regions (from top to bottom: A,B,C and D) and their surroundings, including a bright emission line star located to the northwest of region A. A ridge of emission can be seen, which appears to emanate from region C, and extend up to the northeast. Visible in the box to the upper left is the newly discovered \hii region G0.008-0.07. Right: An enlarged view of the G-0.02-0.07 \hii regions,  the focus of this paper, with a stretch chosen to enhance the various filamentary features visible in the \pa image. Regions A through D are individually labeled, as are several features of interest, including a dark lane separating region A from a slight extension to the northeast, and a series of filaments extending outward from the West side of region A. }
\label{palpha}
\end{figure*}

Regions Sgr A-A, Sgr A-B, and Sgr A-C (hereafter A, B, and C) have similar sizes ($\sim$ 10 arcsec or 0.4 pc in diameter), and exhibit shell-like morphologies in various stages of disruption. Region A is the brightest, as well as the most extended.  It has a semicircular shell shape which appears open on the western edge, where there is an unusual series of roughly parallel linear features which decrease in brightness toward the west of the \hii regions.  The nature of these features is discussed in more detail in Section \ref{A_fil}. On the northeast edge of this region, at approximately the ten o'clock position (see Figure \ref{palpha}, right) there also appears to be a dark lane separating the main shell from a slight extension. We discuss the nature of this feature further in Section \ref{Dlane}. 

Region B is the faintest of the three, and has a complete, albeit faint, shell morphology which can be seen in Figure \ref{palpha}. Its circular shape suggests that it is still embedded on all sides in the natal cloud, or is being viewed from a different angle than A and C. Like regions A and C, the eastern side of this shell appears brightest, thickest, and has the best defined edge. 

Region C has a larger opening angle than region A, and the shell appears discontinuous on the northern and southern edges. The nature of the western edge of  region C is unclear; possibly it is part of the original boundary of the region, or alternatively it may be that the star responsible for ionizing region C is also ionizing another, nearby cloud front. In addition to these prominent features, there is a faint larger-scale ridge of diffuse \pa emission that can be traced in Figure \ref{palpha} from the southeast edge of region C, where its shell appears to end discontinuously, several parsecs toward the northeast. This faint emission appears to trace the surface of the M-0.02-0.07 cloud as seen in 450 micron continuum images from \cite{PP00} (Figure \ref{color}).

\subsection{Region D}
\label{Dshape}

Sgr A-D (hereafter D) is the most compact region. With a size of 0.06 pc (RA) $\times$ 0.2 pc (dec), this region is at the upper end of the distribution of ultra-compact \hii (UC\hii) region sizes, which typically have sizes less than 0.1 pc \citep{Churchwell02}. 

The 8.4 GHz flux we measure for D is very different than that recently measured for this \hii region using the same data set \citep{YZ10}. They find the total flux for both peaks of this region combined to be 60 $\pm$ 4 mJy, whereas we find a flux of 105 $\pm$15 mJy. We are satisfied our measurements of region D do not resolve out any of its flux, and further note that our result is much more consistent with the expected flat spectrum of an \hii region at radio frequencies, given the published 14 GHz and 5 GHz fluxes for this \hii region (see Table \ref{data}). 

The \pa morphology of region D is irregular.There are two bright peaks which are slightly north-south asymmetric, each peak appearing to have a tail of emission extending toward the south. The peaks appear composed of several clumps, although these clumps are near the limits of the spatial resolution of the \pa observations. The two peaks are separated by an apparent void of \pa emission, similar to that seen on the northeast edge of region A. The nature of this void is discussed further in Section \ref{Dlane}.

\subsection{New Radio Sources}

In producing the radio images, we discovered two new radio sources in the field of view (see Figure \ref{radiofig}). One source, G0.008-0.07, is extended in the 8.4 GHz images, and has a morphologically similar counterpart in the \pa data. In the \pa image (Figure \ref{palpha}), the radio source appears to consist of two regions of diffuse emission surrounding several brighter compact knots. The other radio source, G-0.04-0.12, is a faint, compact region of emission which lies just outside the boundary of the area surveyed in \pa. However, it appears to have a faint counterpart in 24 $\mu$m images of the Galactic plane \citep{YZ09}, suggesting the radio emission to be thermal in nature.

\subsection{\hii Region Properties}

As the radio data are not affected by extinction, we use them to determine the physical properties of the \hii regions. Traditionally, HII region parameters are determined by assuming the geometry of  a uniform-density sphere \citep[e.g.,][]{MH67}.  This geometry predicts a peak in emission at the center of the \hii region, and is a good approximation for unresolved \hii regions, or partially-resolved \hii regions such as region D.  

However, the larger \hii regions A, B, and C are resolved and exhibit an edge-brightened morphology inconsistent with this geometry. We found that assuming the geometry of a uniform sphere for these \hii regions inflated their radii, thereby diluting the measured electron density, as well as overestimating the mass of ionized gas. 

We instead determined parameters of the HII regions A, B, and C by modeling them as shells of uniform density, adjusting the outer radius and thickness of the shell to fit the observed HII region profiles. Our method of fitting a shell model is still a rough approximation of the \hii region parameters, in fact, we expect a density gradient through the shell, with the highest density occurring at the ionization front.  However, more detailed models are not warranted by the present data set.

To determine a representative radial profile for each shell, we azimuthally average the intensity of each shell (in Jy/beam) over the angles where the shell is most continuous, which are indicated by the shaded regions in Figure \ref{shell_fit}. This simple model (given by Equation \ref{eqa}) can then be fit to the averaged profile in order to solve for the parameters of each \hii region,  which are reported in Table \ref{data}. The intensity of emission from the shell as a function of position is given by:

\begin{small}
\begin{subequations}
\noindent

	\begin{equation} \textrm{I}_{\nu} = 1680\, \langle\, g_{ff} \rangle\, \textrm{T}_{\textrm{e}} \hspace{0.05cm}^{-0.5} \hspace{0.1cm} \Omega_{\textrm{beam}}	\hspace{0.1cm} \textrm{L(p)}\, n_{\textrm{e}}^2 
	\label{eqa}
	\end{equation}
	\noindent
	\begin{equation} 
	\textrm{L(p)}\hspace{-0.1cm}=2
	\begin{cases}
 		\left(R^2\hspace{-0.1cm}-p^2\right)^{\frac{1}{2}}\hspace{-0.1cm}-\hspace{-0.1cm}\left(\left (R-tR\right)^2\hspace{-0.1cm}-p^2 \right)^{\frac{1}{2}} &\hspace{-0.4cm}, \hspace{0.1cm} p\le R-tR\\
 		\left(R^2\hspace{-0.1cm}-p^2\right)^{\frac{1}{2}}  &\hspace{-0.4cm}, \hspace{0.1cm}p>R-tR
	\end{cases}
	\label{eqb}
\end{equation}

\end{subequations}
\end{small}

Here L(p) is the path length through the HII region at projected offset p from the center, R is the outer radius of the shell, and t is the thickness of the shell as a fraction of R. 

The emission measure is given by L(p)$n_{\textrm{e}}^2$, and the mass of ionized gas in the the \hii region can be calculated by multiplying the RMS electron density by the volume of the shell section used to compute the radial profile (the shaded area in Figure \ref{shell_fit}). The calculated masses only account for the mass of this shaded portion of the \hii region; there is also some emission outside of the modeled area which is not accounted for, and if corrected for would lead to an increase in the total M$_{\textrm{\hii}}$. We can estimate the magnitude of this correction from the percentage of the total \hii region flux outside of the modeled area, which is 17\% of the flux for region A, 4\% for region B, and 30 \% for region C.

The Lyman continuum flux required to ionize each nebula, which yields the spectral type of the star primarily responsible for the ionization, is largely independent of the geometry assumed, depending only upon the distance to the \hii regions and the flux and temperature for each \citep{Rubin68}. The temperatures are taken from the recombination line measurements of \cite{Goss85}. The stellar types we derive for the dominant ionizing source in each nebula (O7-O9) are consistent with those previously determined by \cite{Goss85} and S92, who both concluded the \hii regions were each consistent with being ionized by a single O star.

Although the radio data allow us to determine the spectral type of the stars ionizing each \hii region, the ionizing source of each nebula remains unidentified. Neither A,B, nor C have a detectable associated emission line stellar counterpart in the \pa map. It is also not possible to uniquely identify a central star or stars responsible for ionizing any of these nebulae against the stellar background in the 1.90 $\mu$m continuum images. Each of these \hii regions has several dozen stars inside of its boundary, and if the \hii regions are indeed stellar wind bowshocks, as suggested in a recent interpretation of their kinematic structure \citep{YZ10}, then the primary ionizing source may be offset from the geometric center of the nebula. Near-infrared integral field spectroscopy of these regions \citep{Cotera99} indicates there are three stars which may be potential ionizing sources for regions A,B and C. They have  spectra devoid of strong emission or absorption features, and thus could be main sequence O stars, however no follow-up work has been done to verify their nature. The only emission line star we see in this area in our \pa images is a previously identified Wolf-Rayet star to the northwest of region A \citep[see Figure \ref{palpha};][]{Cotera99}. This star appears not to be related to these \hii regions, as we see no evidence of the expected ionization front were it neighboring the M-0.02-0.07 cloud. 

\section{Extinction from the Paschen $\alpha$ and Radio Continuum Data}

\subsection{Calculating the Extinction}

Together, the 8.4 GHz radio continuum maps and \pa images can be used to determine the extinction toward each \hii region. Although the emission mechanisms are different, the \pa and the radio emission trace the same ionized gas component, and the intensity of both is proportional to the square of the electron density. The free-free radio emission suffers little or no extinction, whereas the \pa emission will be significantly reduced by extinction. By comparing the observed flux density ratio between the radio and the \pa to the theoretical expectation, we can then determine by what factor the \pa emission has been reduced, and thus find the dust extinction at 1.87 $\mu$m toward these regions along the line of sight.

Following the calculations of \cite{Scoville03}, hereafter S03, who similarly derive the extinction from NICMOS \pa observations and 5 GHz continuum observations, we estimate the expected flux per pixel for both \pa and the 8.4 GHz continuum, using Table 4 from \cite{Osterbrock89}.

Assuming case B recombination, and in the event of zero attenuation, the intrinsic flux in the \pa line, per pixel, is given by:

\begin{equation} F_{P \alpha}=6.4\left( \frac{T_e(K)}{6000} \right) ^{-.87}\frac{n_e n_p l a}{4 \pi d^2} \textrm{ mJy Hz}
\label{eq0}
\end{equation}

Here, $n_e$ and $n_p$ are the electron and proton densities, $l$ is the path length in the ionized gas, $a$ is the projected area of a pixel on the sky, d is the distance, and $T_e$ is the electron temperature, values of which have been previously calculated for G-0.02-0.07 using H92$\alpha$ recombination line measurements and assuming LTE \citep[][see Table \ref{data}]{Goss85}.

The flux density per pixel of the radio continuum emission can be similarly expressed:
\begin{small}
\begin{equation} S_{f\!f}=4.2\times10^{-13} \! \left( \frac{T_e(K)}{6000}\right)^{\!\!-.35}\!\left(\frac{\nu}{5 \textrm{GHz}}\right)^{-.1}\frac{n_e n_p l a}{4 \pi d^2} \textrm{ mJy}
\label{eq1}
\end{equation}
\end{small}

All of the input variables here are assumed to have the same values as for the \pa emission, although the temperature dependence is different. As a result, the intrinsic ratio of the \pa line flux to the radio free-free flux density can be expressed as:

\begin{equation} \frac{F_{P \alpha}}{S_{f\!f}}=1.5\times10^{13} \left( \frac{T_e(K)}{6000}\right) ^{\!\!-.52} \left( \frac{\nu}{5 \textrm{GHz}} \right) ^{-.1} \textrm{ Hz}
\label{eq2}
\end{equation}

The factor by which the observed ratio of \pa to 8.4 GHz emission is reduced compared the theoretical expectation yields the 1.87 $\mu$m extinction.

\subsection{The Choice of Extinction Law}

To calculate the extinction in more standard visual magnitudes or A$_{V}$, an extinction law must be assumed. We adopt the near-infrared extinction law of \cite{Nishiyama08}, which has been widely used for recent GC studies. This law is specific to the particular properties of dust and molecular clouds toward the GC and has a $\sim \lambda^{-1.99}$ power law form. Applying this law yields A$_H$/A$_V$ = 0.108 and A$_{Ks}$/A$_V$= 0.062. The \pa line (1.87 $\mu$m) lies between the H (1.6 $\mu$m) and K (2.2 $\mu$m) bands, and we fit a power law equation A$_{\lambda}/A_{V} = 0.29  \lambda^{-1.99}$ to these values to determine A$_{ P \alpha}$/A$_V$ The resulting equation to determine A$_V$ from our 8.4 GHz flux density (S$_{f\!f}$) and \pa flux measurements is as follows:

\begin{equation} A_V = 30.4 \times \textrm{log} \left( \frac{(F_{P \alpha} / S_{f\!f})_{intrinsic}}{(F_{P \alpha} / S_{f\!f})_{observed}} \right)
\label{eq3}
\end{equation}

This law gives significantly different results than the law of \cite{RL85} which was previously the standard for GC work, and was used by, e.g., \cite{Scoville03} in their \pa study of Sgr A West. Adopting the \cite{RL85} law would change the constant in Equation (\ref{eq3}) from 30.4 to 18.1, and thus would make our measured extinctions substantially lower. We discuss the comparison of our results with previous extinctions measured using the \cite{RL85} law further in section \ref{ext}. 

\subsection{Constructing the Extinction Map}

The steps to make an extinction map include aligning the 8.4 GHz radio and \pa maps, matching the pixelization (0\arcsec .15 pixels) and smoothing the \pa map to the resolution of the radio CLEAN beam ($1''.85 \times 0.6''$) with AIPS tasks HGEOM and CONVL. In the process we found that the \pa and 8.4 GHz images were offset by 1''.2 in right ascension, and so after smoothing the \pa image we performed a normalized cross-correlation of the diffuse emission in the two images in IDL with the procedure CORREL\_OPTIMIZE to find the translation that resulted in optimal alignment.  Even though the \pa may be affected by non-uniform local extinction, we still expect that the radio and \pa images will trace the same structure in these \hii regions, and so this should yield the proper alignment of the two maps. There are no point sources other than Sgr A* in the radio image to compare, however the positions of stars in the \pa survey images with known SiO masers  have been compared to the catalog of \cite{Reid07}, and the astrometric uncertainty for all the \pa images is measured to be $\sim$ 0.05\arcsec (Dong et al. 2011). This suggests that the majority of the offset originates in the radio reference frame. The radio frame was then corrected to match the \pa data, the  \pa image was divided by the radio image, and Equation (\ref{eq3}) was applied so that the pixel values represent the local value of A$_V$. 

Before constructing the final map, the radio map was also clipped to the 3 $\sigma$ level to eliminate any extraneous peaks in the background noise. Other apparent extinction peaks may still occur where there are over-subtracted stars in the \pa images. 

The extinction has a slight dependence on temperature (Equation \ref{eq2}), and as temperatures for each \hii region were measured by \cite{Goss85}, this was taken into account for the extinction values we calculate for each individual \hii region. However, this effect is small: for example, lowering the temperature of region D by 2000 K results in only a $4\%$ change in the median extinction measured for that source.  

\subsection{ Measured Extinctions}

The extinction maps which we derive, shown in Figure \ref{AV}, are the first measurements of the extinction structure across these regions.  These maps are, however, limited in that they only contain information for areas with emission from ionized gas. Unphysical extinction values could also result from nonthermal radio emission, such as that from the supernova remnant Sgr A East. However, Sgr A East is sufficiently separated from the \hii regions that it does not appear in our figures, and should not bias the results presented here. Due to the substantial difference in the visual extinctions calculated using different extinction laws, we report in Table \ref{extinction} the 1.87 micron extinction, which is not affected by the choice of extinction law, in addition to the A$_{V}$ we calculate using the \cite{Nishiyama08} law. To measure the extinction toward each \hii region, we calculate the median of all pixels in our map from each \hii region. We report the median extinction toward each individual \hii region in Table \ref{extinction}. 

\subsubsection{ Regions A, B, and C}

The median extinction values measured for regions A,B, and C are all around A$_{V}$=44.6, and the maximum values are similar as well, about A$_{V}$ = 52.  Some parts of these regions, such as the westernmost ionized ridge of region A, and the diffuse interior of region C, are too faint to appear above the noise level in the 8.4 GHz map, and so
there is no extinction information for these areas. The extinction measured across each \hii region is relatively uniform, varying locally by 3-4 magnitudes. We also observe that the maxima in extinction for regions A and B are located near their apparent mutual boundary, on the southern edge of region A, and on the northern edge of region B. 

The center of the 8.4 GHz image of region C is somewhat affected by the previously mentioned background striping artifacts of the radio data, leading to enhanced emission in its center which appears as a slight peak in the extinction map in this region. As a result, the extinction values at the center of region C should not be considered reliable.

\subsubsection{ Region D}

As region D is mostly unresolved at the resolution of our extinction map, we report only the maximum values of the extinction toward D1 (the eastern peak of the \hii region) and D2 (the western peak). These values are A$_{V}$ = 69.1 and A$_{V}$ = 70.5, respectively, which is almost 20 magnitudes additional extinction than the maximum extinction measured for regions A, B and C.  Comparing the extinction map of region D to the radio and \pa maps (Figure \ref{Ddetail}),  we see that the eastern peak in the extinction map is slightly offset from the peak of D1 in the \pa emission. We measure the area between D1 and D2 to have a minimum extinction of  A$_{V}$ = 65.6. However, as the two peaks of region D are not fully resolved in the radio image or the resulting extinction map, this extinction value is likely not representative of what appears to be a void in the ionized gas emission in both the radio and \pa images, and instead most likely results from the overlapping PSFs of the two point-like peaks D1 and D2. On the northern edge of region D is a diffuse extension that appears only in the 8.4 GHz images, likely because it is too faint or extinguished to appear in Paschen alpha. We find the lower limit of the extinction toward this structure to be A$_{V}$ = 60.7 magnitudes.

\section{ Discussion}

\subsection{Comparison with existing extinction results}
\label{ext}
Average extinctions have been previously measured for the individual \hii regions in the Sgr A East complex at 12.8 microns (S92), and using Brackett-$\gamma$ \citep[hereafter C00]{Cotera00}. S92 derived approximate extinction values for each of the \hii regions at 12.8 $\mu$m from fractional ionic abundances of Ne, S, and Ar measured from mid-infared fine structure lines.
Our results are consistent with their findings that region D suffers substantially higher extinction than the other three regions.  They conclude that regions A-C are located at the front edge of the cloud, with D more embedded. Although our results are qualitatively the same, it is difficult to more quantitatively compare our values. The recently determined GC extinction law of \cite{Nishiyama09} covers infrared wavelengths up to 8.0 $\mu$m, and although it shows the mid-IR extinction law to be quite flat, the extinction curve immediately beyond 8 $\mu$m is known to rise sharply due to the wide 10 $\mu$m silicon bump.  Using the Nishiyama law, the median A$_V$ values measured toward A,B, and C ( all $\sim$ 44.6 magnitudes) correspond to 8 $\mu$m extinctions that are comparable to the 12.8 $\mu$m extinctions estimated by S92 (see Table \ref{extinction}). However, the peak extinction we measure toward region D corresponds to an 8 $\mu$m extinction substantially less than that S92 report at 12.8 $\mu$m. 

Our \pa\hspace{-2pt}-derived extinctions are consistent with extinctions calculated in a similar manner by C00 using Br $\gamma$ imaging along with 6 cm data from \cite{YZM87}.  Using J, H, and K' colors, C00 also determined a median stellar extinction (hereafter MSE) along the line-of-sight toward the \hii regions with 1\arcmin\, resolution. They found that the MSE was consistently higher than the \hii region extinctions, leaving open the possibility that regions A, B, and C are located in the foreground of the GC. However, it is important to remember that the MSE and the \hii region extinctions can sample a fundamentally different volume.  The Br $\gamma$/radio extinction measures all of the extinction along the direct line of sight to a given \hii region while the MSE, in contrast, is derived from the colors of stars which may be distributed in front, behind, or around the \hii region in question. Thus, if the cloud is not totally opaque (and therefore some stars with large extinctions can be observed behind the cloud), the MSE can be larger than the extinction measured toward the \hii regions even if the HII regions are also located at the GC .

We also compare the extinctions we derive with the extinction map of \cite{Schul09}, which is the most recent large-scale map (2\arcdeg $\times$1.4\arcdeg) of GC MSE,  constructed using Spitzer-IRAC mid-infrared colors of long-period variables. This map, however, has very low resolution, with pixel sizes of 2\arcmin\, on a side. In the four pixels of the map which overlap the \hii regions, \cite{Schul09} measure extinctions of A$_{V}$ = 26,32,46, and 48.  As these pixels are very large compared to the size of the \hii regions and even compared to the M-0.02-0.07 cloud, we interpret these values as global averages, biased by the filling fraction of cloud in each pixel. Even in the two pixels which have the highest extinction and overlap the largest portions of the molecular cloud, the measured extinction values likely significantly underestimate the extinctions present in the small-scale substructure or densest cores of the cloud. We thus interpret our extinction values as consistent with regions A-C being at a similar distance as this cloud, though likely in front of it due to the uniformity of extinction across the three regions, and region D being embedded in an especially dense core of the cloud. 

The median extinction values we found for regions A-C (A$_{V} = 44.6$) correspond to extinctions of $A_{V}$ = 26, if we use the extinction law of \cite{RL85}.  This is similar, though slightly lower than the extinction values measured by  \cite{Scoville03} for Sgr A West using the same law, which vary from A$_{V}$ = 20 to 50, with a median value of A$_{V} \sim$ 31. As Sgr A West is not significantly occulted by either the M-0.02-0.07 or M-0.13-0.08 molecular clouds, its extinction should be due as well to the foreground screen from intervening spiral arms. The higher median extinction observed in that direction, relative to that observed toward the G-0.02-0.07 \hii regions, may be partly due to the fact that Sgr A West suffers somewhat higher extinction on its periphery due to the surrounding circumnuclear disk of molecular gas \citep{Scoville03}, which likely biases the median value upward.

In summary, our results agree with those of S92, confirming that the extinction toward A-C is consistent with a location at the GC, but is sufficiently low and uniform toward these regions that it is not consistent with significant local attenuation from M-0.02-0.07. The higher and non-uniform extinction of D, in contrast, suggests it is embedded in a dense core of the M-0.02-0.07 cloud.

\subsection{The Nature of Region D}
\label{Dlane}

\cite{YZ10} recently examined the kinematic structure of region D using mid-infrared spectroscopy of the Ne II line. They found that the two peaks of region D have very different velocities, with the eastern peak appearing redshifted, and the western peak appearing blueshifted, both by $\sim 30$ km s$^{-1}$ with respect to the ambient cloud velocity. The authors argue that this kinematic structure, as well as the observed width of the Ne II line emission, are best described by a collimated outflow or jet from the central star which is impacting a surrounding, evacuated cavity. Based on the observed velocity shifts, the western edge of the disk is tipped toward us, with an estimated disk position angle of $70\arcdeg$.  

Our observations largely support this model, though we disagree on a few points.  We interpret the continuum source seen in the 1.87 and 1.90 $\mu$m images (Figure \ref{D187190}) as more likely to be a star than continuum emission from the western peak of region D or scattered light. The source was found by \cite{Cotera99} to have a stellar spectrum and an H-K' color of 3.5, which with our adopted extinction law corresponds roughly to an extinction of A$_{V}$ = 76, very similar to the peak extinction of region D. It is thus very likely that this star is the source of the ionization of region D.  The ionizing source for region D is then much more offset from the centroid of the \hii region than implied by the model of \cite{YZ10} in their Figure 9, lying just to the southeast of the western \pa peak (our Figure \ref{Ddetail}, Left). 

\cite{Cotera99} also suggest that this star is a B[e] type star based on the detection of weak He I and Br $\gamma$ line emission in its spectrum. With the higher spatial resolution of the \pa data, we do not resolve significant line emission coming from the stellar point source (see Figure \ref{Ddetail}, Left). It is likely that the spectrum from \cite{Cotera99} resulted from a superposition of nebular emission lines from the UC\hii region and the stellar continuum from this star.

The strongest evidence for a disk in region D from our data is the apparent void of radio and \pa emission between the the two peaks of region D , a dark lane running north-south through its center in both radio and \pa images. We suggest that this dark lane represents a region of largely neutral gas which has been shielded from the ionizing radiation of the central star by the disk (see Figure \ref{Dmodel}). Absent a disk, one might expect to see a more continuous spherical shell of ionization around the star.  Although our extinction map does not clearly identify a peak of extinction at or around the ionizing star corresponding to this disk, this is not inconsistent with the presence of a dense disk. We can explain the lack of significant extinction detected toward this disk if the disk is small, and thus unresolved by our observations, or if the disk does not occult significant ionized emission along the line of sight, in which case we would have no information on the extinction toward the neutral gas, including the disk, along the line of sight of the disk. An example sightline for which this would be the case is shown in Figure \ref{Dmodel}.

We concur with \cite{YZ10} that the asymmetry of the emission from region D, with the blueshifted emission arising much closer to the central star than the redshifted emission, is almost certainly due to the \hii region being embedded in a density gradient (see Figure \ref{Dmodel}). The extinction measured toward the eastern peak of D is slightly lower than that measured toward the western peak, and the \pa emission falls off much more steeply on the western edge of  D, suggesting a more steeply increasing column density gradient in that direction.  This increase in column density corresponds to the location of the dense western ridge of the M-0.02-0.07 cloud, located between the \hii regions and Sgr A East (see Figure \ref{color}).  Indeed, a comparison of the position of region D to higher resolution observations of dense gas traced by ammonia (1,1) emission in the western ridge \citep{Coil00} shows that region D appears to lie on the eastern edge of a dense core (see Figure \ref{nh3}). Based on its extinction, region D is likely embedded in or behind this core.

A structure reminiscent of region D is also seen on the northeast edge of region A (Figure \ref{A_lane}). A protrusion of emission is separated from the main shell of region A by another apparently dark lane, exhibiting a lack of emission in both \pa and 8.4 GHz images.  This protrusion of emission is also resolved in the Ne II spectra of \cite{YZ10}, but appears not to have the same kinematic structure as region D: emission on each side of the dark lane appears to have the same radial velocity. It is still possible that this structure, like region D, is a young massive disk, with the dark lane corresponding to the shadow of the disk, but either it has no collimated outflows, or we are observing this system closer to edge-on. Like region D, a star is visible slightly offset from the center of the dark lane. The star is visible in near-infrared images of \cite{Cotera00} and appears similar in color to the star in region D, though no value for its H-K' color is reported. 

To verify the presence of a disk in regions A and D, one could observe these regions at high spatial-resolution in the millimeter and radio regimes to search for warm dust or molecular gas in the disk, or even free-free emission from the surface of the disk. Higher resolution spectra of the stars in regions A and D could also help determine whether their properties are consistent with extremely young, massive stars. 

\subsection{The Ionized Ridges of Region A}
\label{A_fil}

To the southwest of region A lie three roughly linear ionized ridges with increasing separation from the opening of the \hii region (Figures \ref{palpha}, \ref{A_fils}). While it is possible that these ridges could be pre-existing structures that are being illuminated as the central star of region A passes nearby, their unusual alignment with each other and with the opening of region A suggests a closer relationship. We interpret them as most likely to be the interaction between an ionized flow from inside region A and the diffuse surrounding ISM. These limb-brightened shells would propagate outward at the sound speed. If, as suggested by \cite{YZ10}, region A is moving both to the east and toward us, then we should be able to see a difference in velocity between the radial velocities of region A and the expanding shells. However, the magnitude of such a velocity difference would be the sound speed ($\sim$ 10 km s$^{-1}$) projected along the line of sight, and for motion 20 to 30 degrees out of the plane of the sky would correspond to a velocity difference of only 3 to 5 km s$^{-1}$.  In Figure 8 (Left) of \cite{YZ10}, two of the ridges are seen to have radial velocities around $\sim 50$ km s$^{-1}$, similar to the mean radial velocity of the \hii region, and of the ambient medium of  the M-0.02-0.07 cloud. The data show no evidence for a velocity difference between region A and the ridges of greater than 5 km s$^{-1}$, but the data lack the velocity resolution to conclusively determine whether a smaller velocity shift is present.   

Alternatively, it is possible these structures could be due to an instability first proposed for the case of old planetary nebulae passing through a magnetized ISM, essentially a magnetic Rayleigh-Taylor instability \citep{Dgani98,Soker97}.  As post-shock material cools isothermally, it is subject to a magnetic Rayleigh-Taylor instability, stabilized in the direction perpendicular to the magnetic field, leading to a density pattern of ridges in the ISM behind the \hii region, parallel to the ambient magnetic field. Although the inferred velocity of the central star of region A \citep[30 km s$^{-1}$,][]{YZ10} is slower than the minimum estimated by \cite{Soker97} for significant instabilities to develop (40 km s$^{-1}$), the warm, dense ISM of the GC combined with the likely presence of a strong pervasive magnetic field \citep[$50$ - $100\hspace{0.1cm}\mu$Gauss,][]{Crocker10}, are both factors that should be conducive to the development of such instabilities. However, it is unclear why similar instabilities would not also be seen to be associated with regions B and C, which are believed to be have the same stellar wind bowshock kinematics \citep{YZ10}. 

Higher spectral-resolution observations as well as more sensitive observations of the faint ionized gas in these ridgelike features are necessary to determine whether the ridges of region A have kinematics consistent with shells propagating outward at the sound speed. In addition, if the ionizing sources for these \hii regions were identified, measuring the radial velocities of the stars could test the possibility that the ridges are due to a magnetohydrodynamic instability. 

\subsection{Locating the \hii Regions}

\begin{figure*}
\epsscale{1.0}
\plotone{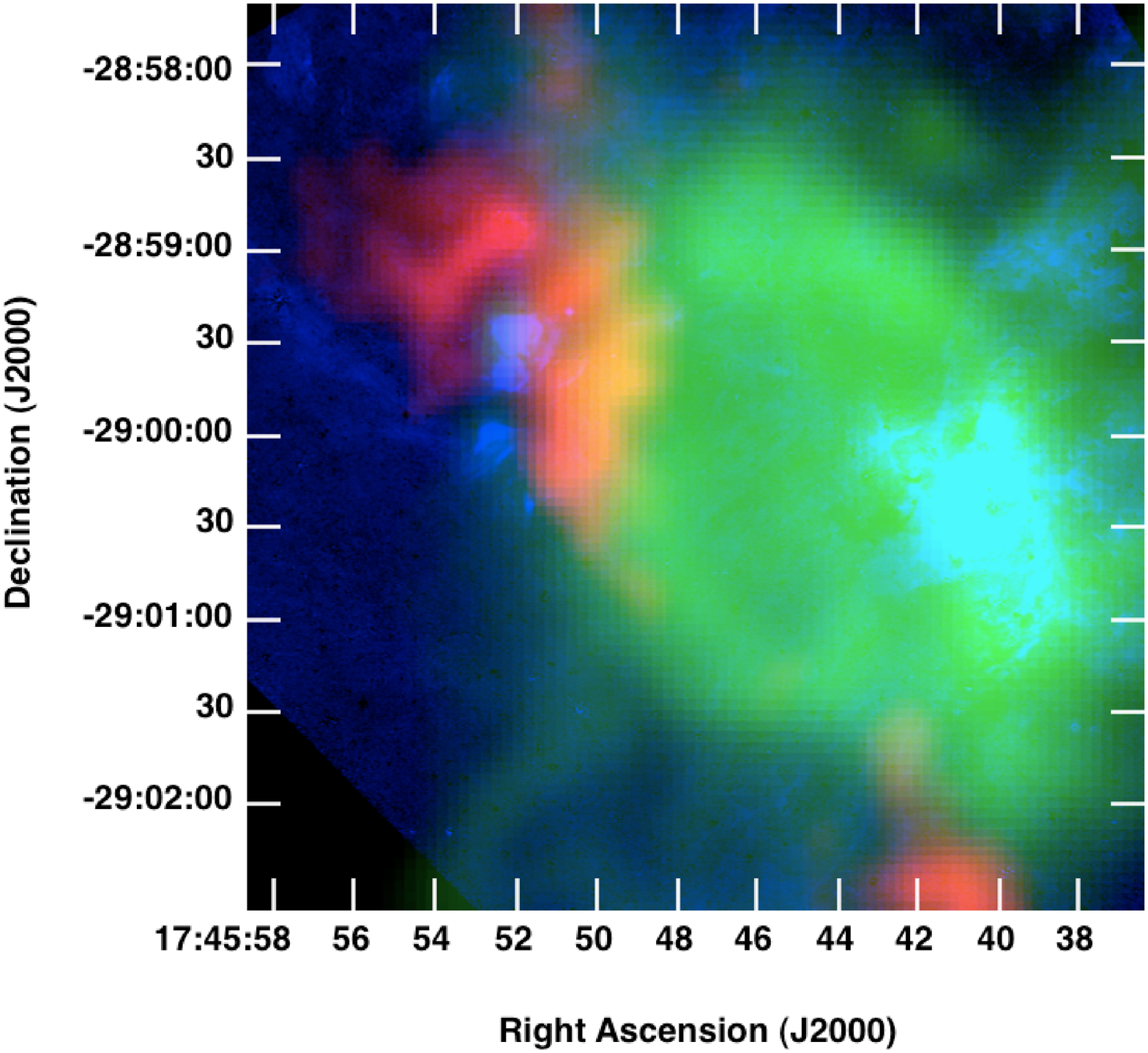}
\caption{Composite image of \pa in blue, VLA 1.4 GHz data in green \citep{Lang10}, and SCUBA 450 microns \citep{PP00} in red. The main structures traced by \pa emission (blue) are the G-0.02-0.07 \hii regions at middle left, and Sgr A West at lower right. The 1.4 GHz image (green) also traces these structures, but most prominent at 1.4 GHz is Sgr A East, the large shell of nonthermal emission that borders G-0.02-0.07 and appears to enclose Sgr A West. The 450 micron image (red) traces an entirely different component, the dense cool dust that corresponds to some of the densest parts of M-0.02-0.07, the 50 km/s molecular cloud, in which the \hii regions appear embedded. At the bottom of the image, part of M-0.13-0.08, the 20 km/s cloud, can also be seen.}
\label{color}
\end{figure*}

\begin{figure*}
\epsscale{1.0}
\plotone{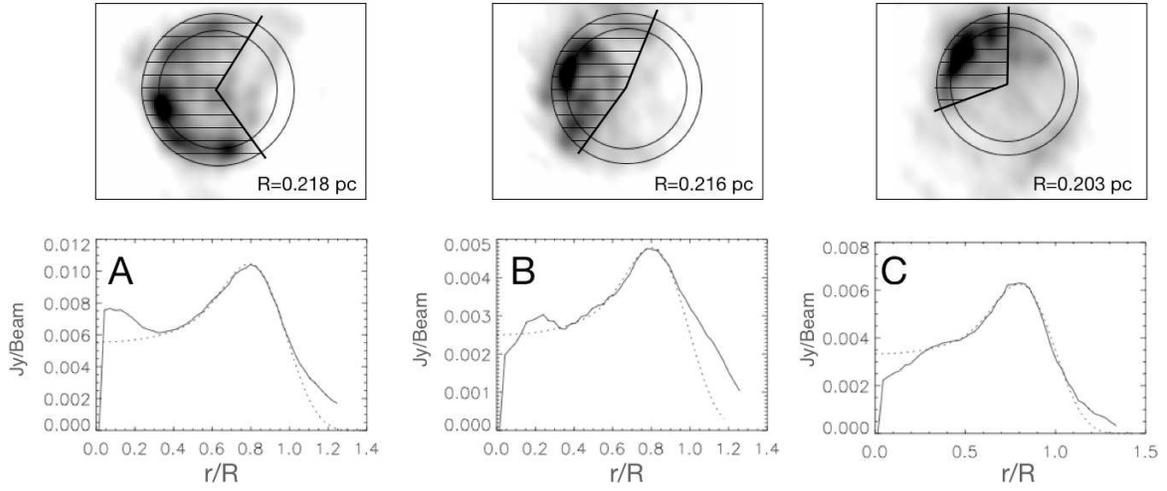}
\caption{ Top: Projected shells (with outer radius size indicated for each shell) are superposed on the 8.44 GHz images used to fit shell parameters. Bottom: Azimuthally-averaged radial profiles for each HII region are shown (solid lines). A shell model of the emission measure (dotted lines) is fit to each HII region, from which the electron density can be derived. }
\label{shell_fit}
\end{figure*}

\begin{figure*}
\epsscale{1.0}
\plotone{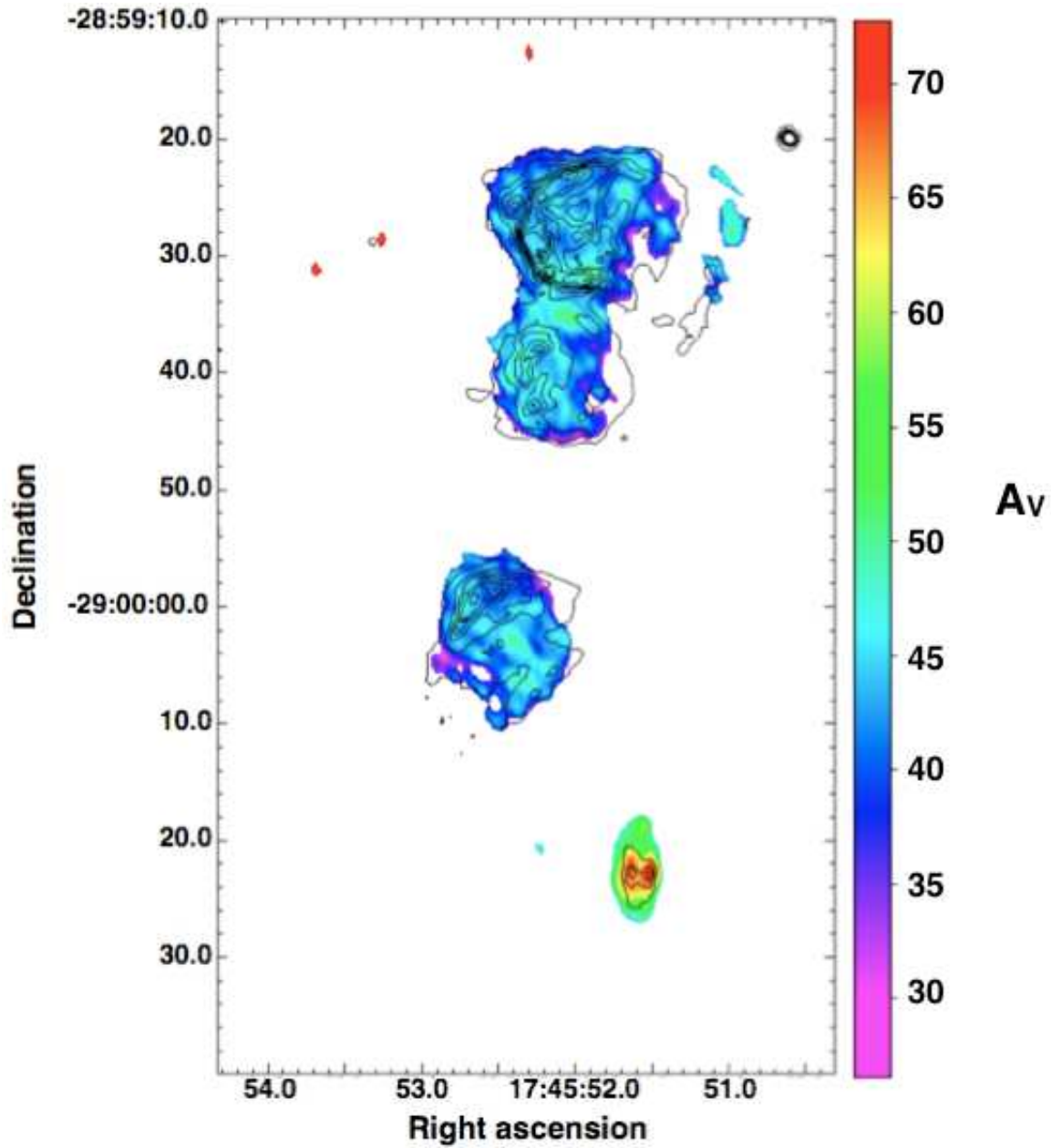}
\caption{ Map of the extinction toward the \hii regions G-0.02-0.07 derived from a comparison between \pa and 8.4 GHz continuum data. The resolution of the map is the same as the 8.4 GHz data, or 1.5 x 0.8 arcseconds. The contours represent the flux density of the smoothed \pa emission. The lowest contour has a values of 0.5 mJy per arcsec$^{-2}$, and the contours are spaced by 0.8 mJy per arcsec$^{-2}$.}
\label{AV}
\end{figure*}

\begin{figure*}
\epsscale{1.0}
\plotone{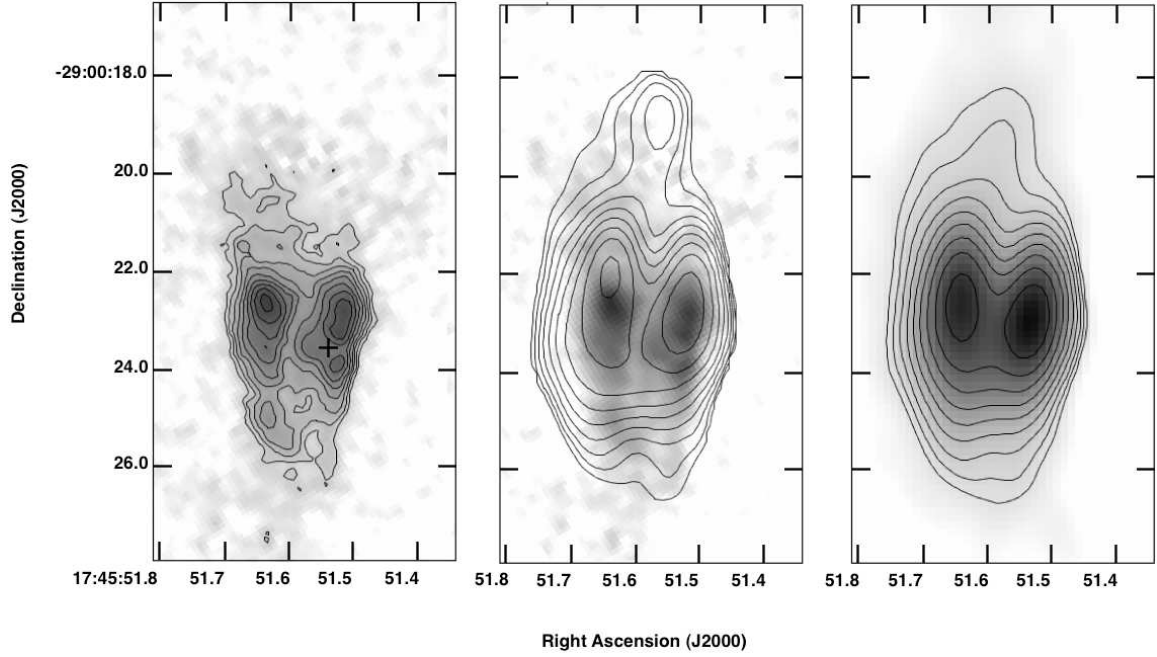}
\caption{ Detail maps of the structure of region D in pure Paschen alpha line emission, with contours of  flux density (Left) and extinction (Center) overlaid. For comparison, the 8.4 GHz image of region D is also shown (Right).  On the Left, the flux density contours have a logarithmic spacing, with the lowest contour having a value of 0.5 mJy per arcsec$^{-2}$, and the highest having a value of 3 mJy per arcsec$^{-2}$. The cross marks the position of the star seen in Figure \ref{D187190}. In the Center, the lowest and highest extinction contours correspond to values of A$_{V}$ = 50 and 70, respectively, with a contour spacing of 2.5 magnitudes. On the Right, the contours on the radio image are logarithmically space, with the lowest contour corresponding to a flux density of 0.5 mJy and the highest corresponding to 20 mJy.  }
\label{Ddetail}
\end{figure*}

\begin{figure*}
\epsscale{1.0}
\plotone{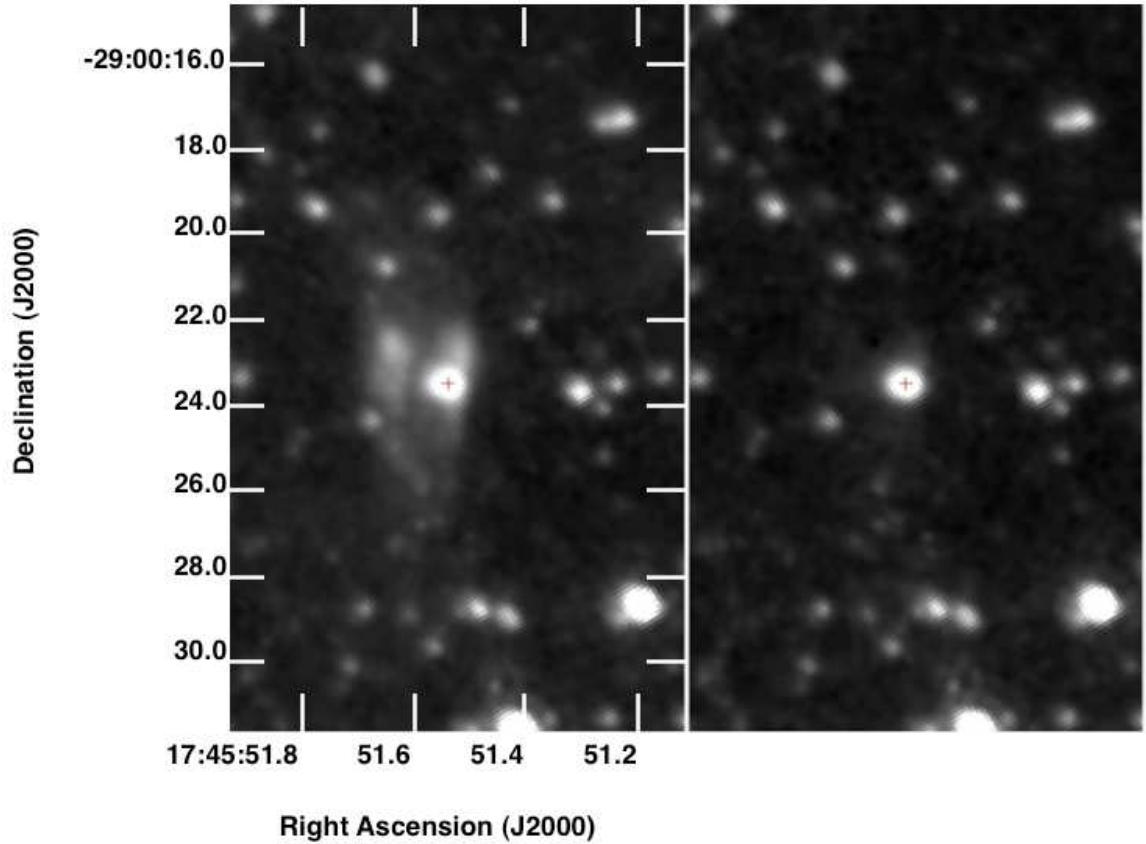}
\caption{ A comparison of the emission toward region D from the F187N and F190N filters. Slightly offset from the center of region D is a point source ( indicated in both images with a cross) which has a F187N/F190N ratio consistent with being purely stellar continuum, with no excess emission from the \pa line. This point source was identified by \cite{Cotera99} as having a stellar spectrum, and classified as a  B[e] star (although this is not confirmed by our lack of detection of a \pa excess). They found the star to be highly reddened (H-K' = 3.5), which corresponds to an extinction consistent with that which we measure toward region D, consistent with it being embedded in this region. }
\label{D187190}
\end{figure*}

\begin{figure*}
\epsscale{1.0}
\plotone{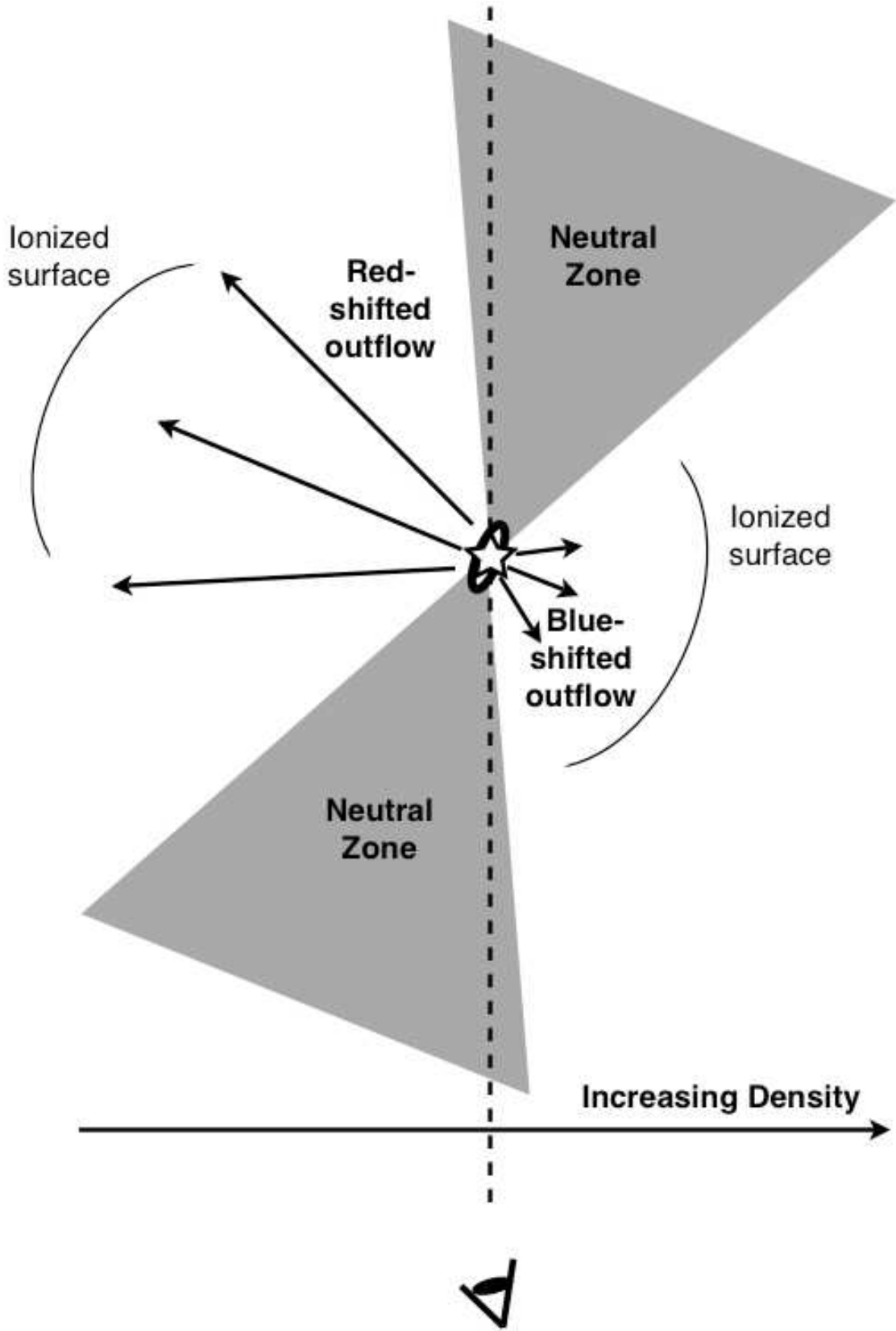}
\caption{ A model of the disk structure of region D.  In this model, the disk at the center is small, and does not occult ionized emission along the line of sight, consistent with the lack of a high extinction value measured toward the location of the disk. We explain the dark lane visible in radio and \pa images which divides the two peaks of region D as due to disk absorption of ionizing radiation from the star. The dark lane represents a region of largely neutral gas which has been shielded from the ionizing radiation of the central star by the disk.  The dominant source of the radio and \pa emission is the ionized surface of cavities evacuated by an outflow from the central disk. The edge of the western cavity appears to be located closer to the ionizing star, suggesting this cavity is smaller. This can be explained if region D is embedded in a density gradient which is increasing to the west. }
\label{Dmodel}
\end{figure*}

\begin{figure*}
\epsscale{1.0}
\plotone{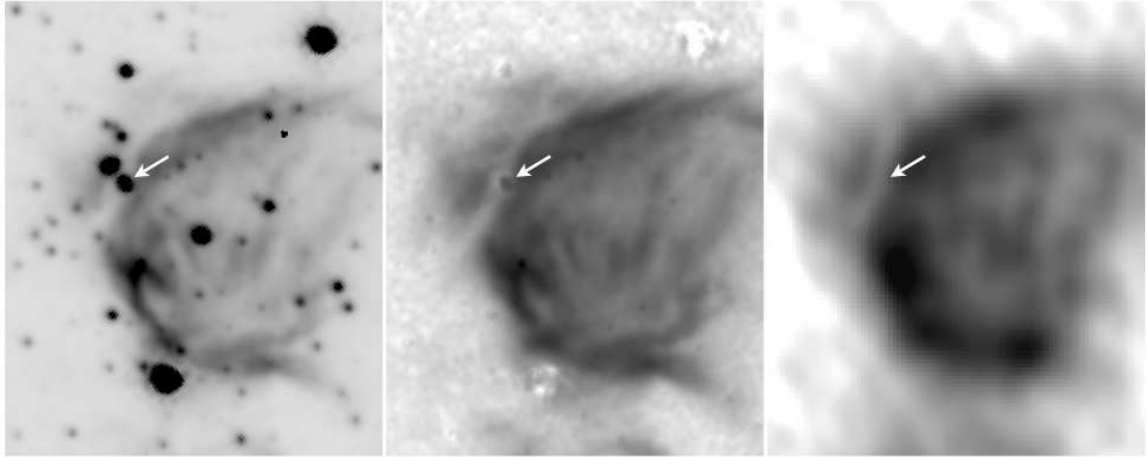}
\caption{ A comparison of the emission toward the northeast extension of region A from the F187N filter, the pure \pa line image, and the 8.4 GHz radio image, from left to right. An emission-free lane is seen separating the main shell of region A from a slight extension at the 10 o'clock position. Like region D, there is a star apparently located slightly offset from the center of this lane ( its location indicated in all images with an arrow). The apparent \pa excess at the position of this star is at has the same intensity as the emission from the surrounding nebula, and is likely an artifact due to the difficulty of subtracting stellar emission in the presence of a diffuse background. No counterpart to the apparent \pa excess at the position of this star is seen in the radio image. In images of \cite{Cotera00}, this star appears highly reddened, but its H-K' color is not given. }
\label{A_lane}
\end{figure*}

\begin{figure*}
\epsscale{1.0}
\plotone{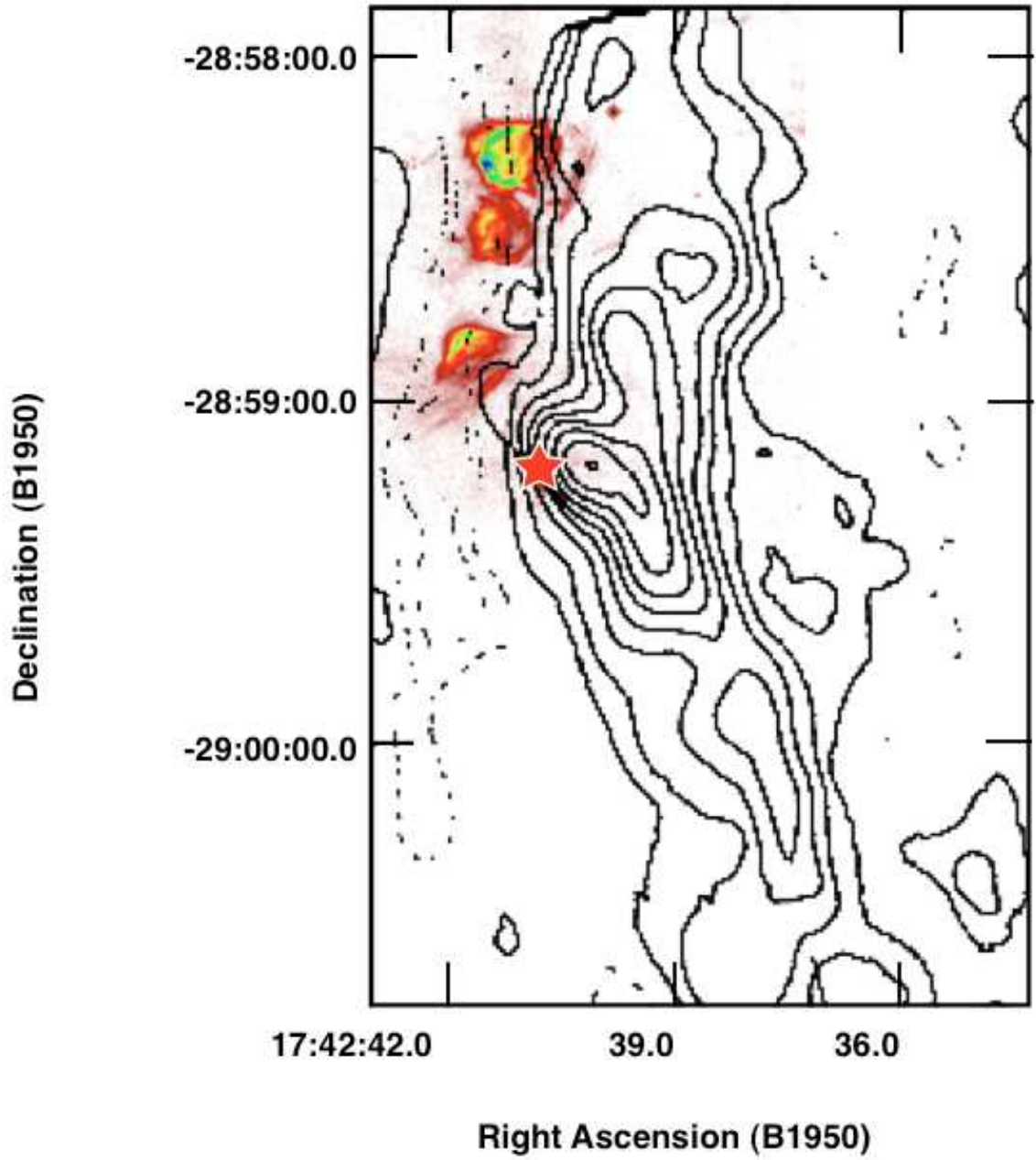}
\caption{ Superposition of Ammonia (1,1) contours \citep[magnification of their Figure 6, Left]{Coil00} and \pa emission, showing the apparent association of region D (indicated with a star) with a dense core in the western ridge of M-0.02-0.07. Region D appears to be immediately adjacent to a steep gradient in the ammonia emission from the core.}
\label{nh3}
\end{figure*}

\begin{figure*}
\epsscale{1.0}
\plotone{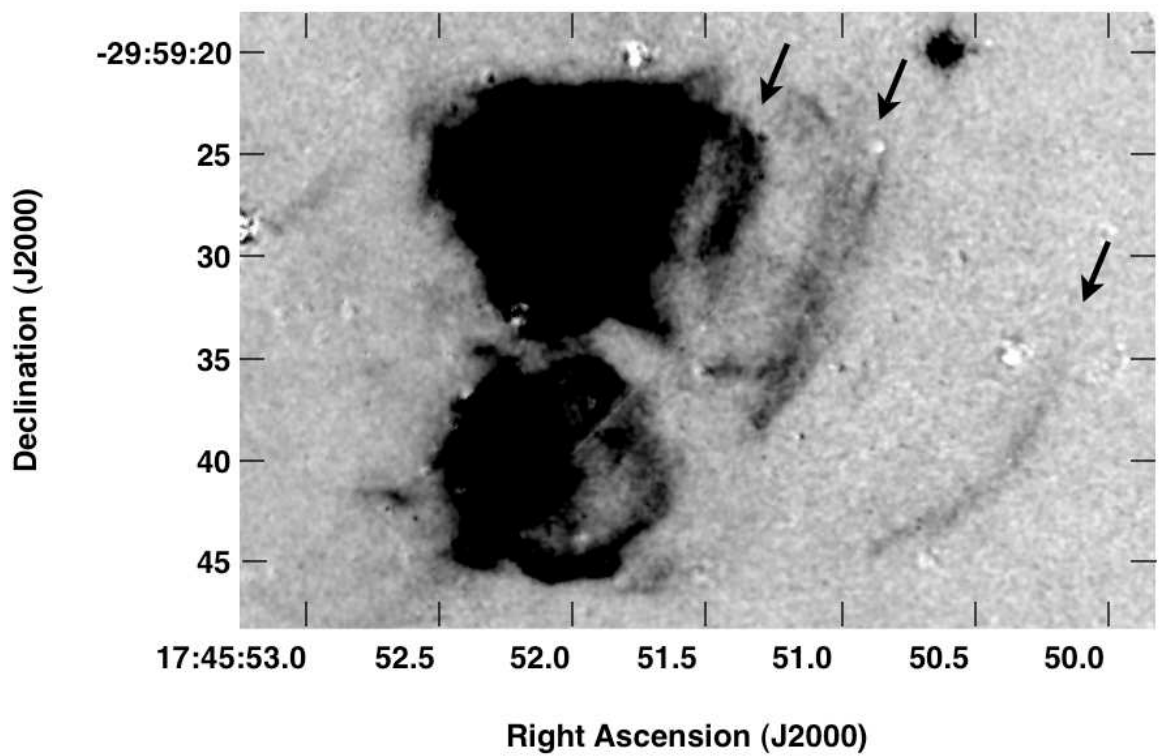}
\caption{ A magnification of the ionized ridges lying to the southwest of region A. The faint line that intersects region B at a 45 degree angle is an artifact from the edge of one of the mosaicked subimages used to construct this map. }
\label{A_fils}
\end{figure*}

Measurements of  consistent radial velocities for the  G-0.02-0.07 \hii region complex and the M-0.02-0.07 cloud (S92) have established that the two structures are associated. However, it is less clear where in the cloud the \hii regions lie. From their CS maps of dense gas in the cloud, S92 divide it into two main structures: an eastern quiescent lobe, lying predominantly to the north of G-0.02-0.07 (the eastern region of 450 $\mu$m emission in Figure \ref{color}), and a western dense ridge (Figures \ref{color} and \ref{nh3}) of gas showing evidence for strong large scale shocks traced by vibrationally excited H$_2$ \citep{YZ01} and 1720 MHz OH masers \citep{YZ96}  resulting from an apparent interaction with the supernova remnant Sgr A East (Figure \ref{color}, in green). The \hii regions appear to lie between the two cloud structures, but their arrangement follows the western ridge, and the eastern periphery of Sgr A East. 

The \pa images and extinction maps presented in this paper provide some additional insight into the location of the individual \hii regions in G-0.02-0.07 with respect to Sgr A East and the two components of M-0.02-0.07. Previous extinction measurements for region A indicated that it was in front of M-0.02-0.07; however, as seen in Figures \ref{color} and \ref{nh3}, the main body of this \hii region and the dense gas in this cloud do not significantly overlap, which suggests the extinction for the bulk of this region would be low even if behind the western ridge. Our extinction map, which provides extinction values for the first time for the faint ionized ridges to the west of region A, shows the extinctions of those ionized ridges to be consistent with that for the rest of the region A, and does not show any evidence for an extinction gradient across the region, as might be expected if the ionized ridges were behind the western ridge of M-0.02-0.07. Assuming these ridges are associated with region A, as their morphology suggests, this places region A in front of the western ridge of M-0.02-0.07. As observations of OH absorption indicate that the western ridge of M-0.02-0.07 lies in from of Sgr A East \citep{Karl03}, region A must also then lie in front of Sgr A East. 

Given the high extinction toward region D, and radio properties and compact morphology which are all consistent with a young UC\hii region still embedded in its natal cloud, we believe that region D is located in the western ridge of M-0.02-0.07. This is consistent with the previously noted apparent close association between region D and a peak in ammonia (1,1) emission in the ridge. If region D is embedded in the western ridge, we would expect to see OH absorption from the cloud against the continuum emission from this region, as is seen where M-0.02-0.07 lies in front of Sgr A East \citep{Karl03}. However, \cite{Karl03} report no absorption for any of the G-0.02-0.07 regions. Examining their Figure 5, however, we do see evidence for some weak absorption (at 10-20 mJy or the 2-4 $\sigma$ level) toward the location of region D in velocity channels 32.4 and 41.2 km/s. As region D is unresolved by the observations of \cite{Karl03}, the absorption may also be somewhat diluted over the beam size. Furthermore, the \hii regions appear to lie in an oversubtracted bowl feature in these maps, which may lead to missing flux for region D. The combination of these effects may explain the relatively weak OH absorption measured by \cite{Karl03} toward region D, suggesting that the OH measurements may not be inconsistent with our finding that region D is embedded in or behind M-0.02-0.07. 

\section{Summary}

We have presented new high-resolution maps of the G-0.02-0.07 \hii regions in the \pa line and have produced extinction maps of these regions using a combination of the \pa maps and archival radio data. The morphologies of these regions and the extinction we measure toward them confirm that they are located in front of, but near to, the M-0.02-0.07 cloud, with region D likely embedded in the dense western ridge of this cloud. In addition, we find that the uniform extinction across region A requires it to be entirely in front of the dense western ridge of the M-0.02-0.07 cloud.

We interpret the series of ionized ridges located to the west of region A as most likely to be a succession of limb-brightened shells resulting from shocks produced as a thermal wind from the HII region interacts with diffuse, ambient gas .

Region D is interpreted as containing a small, opaque disk which shields the neutral gas from ionization by the central star, forming a dark lane which appears to bisect the \hii region in both radio and \pa images. We explain the lack of a measured extinction maximum in this dark lane as being most likely due to the absence of ionized gas along the line of sight of this dark lane.

\section{Acknowledgements}
Support for program HST-GO-11120 was provided by NASA through a grant from the Space Telescope Science Institute, which is operated by the Association of Universities for Research in Astronomy, Inc., under NASA contract NAS 5-26555. This material is based upon work supported by the National Science Foundation under Grant No. 0907934

\bibliographystyle{apj}
\bibliography{G-0.02-0.07final.bib}
\clearpage

\begin{table}[ht]
\caption{\hii Region Parameters} 
\centering
\begin{tabular}{rcccc}
\\[0.5ex]
\hline\hline
& & & & \\
Region & A & B & C & D\\ [0.5ex]
\hline
& & & & \\
8.4 GHz Flux Density (mJy) & $550\pm 75$ & $175 \pm 44$ & $180 \pm 50$ & $105\pm 15$  \\
Radius (pc) & 0.22 & 0.22 & 0.20 & 0.03\\
Electron Density (cm$^{-3}$) & 5100 & 3600 & 4100 & 21000\\ 
Emission Measure (pc cm$^{-6}$) & $6.8e6$ & $3.4e6$ & $4.0e6$ & $3.1e7$ \\
Mass in \hii (M$ _{\odot}$) & 2.2 & 0.8 & 0.4 & 0.1 \\
%Excitation Parameter (pc cm$^{-2}$) & 45.9 & 31.8 & 31.4 &  27.1\\ 
Lyman Continuum Flux ( s$^{-1}$)& $4.8e48$ & $1.3e48$ & $1.6e48$ & $8.1e47$\\
Ionizing Spectrum\footnotemark[1]  & O7 & O8.5 & O8.5 & O9 \\
& & & & \\
\hline
& & & &\\
Temperature\footnotemark[2] (K)& 5800 & 6600 & 5300 & 7400  \\
14.7 GHz Flux Density\footnotemark[2]  (mJy) &  $570\pm 20$ & $173 \pm 20$ & $245 \pm 20$ & $95 \pm 15$  \\
5 GHz Flux Density\footnotemark[3]  (mJy) & $812\pm 100$ & $166\pm 40$ & $196 \pm 40$ & $130\pm 15$  \\
1.5 GHz Flux Density\footnotemark[3]  (mJy) &  $380\pm 80$ & $103 \pm 17$ & $108 \pm 20$ & $45 \pm 7$  \\
& & & & \\
\hline

\footnotetext[1]{ Estimated from \cite{Martins05}}
\footnotetext[2]{ From \cite{Goss85}}
\footnotetext[3]{ From \cite{Ekers83}}

\end{tabular}
\label{data}
\end{table}

\begin{table}[ht]
\caption{ \hii Region Extinctions} 
\centering
\begin{tabular}{cccccc}
\\[0.5ex]
\hline\hline
& & & & &\\
Region & {\bf A} & {\bf B} & {\bf C} & {\bf D1} & {\bf D2}\\ [0.5ex]
\hline
& & & & &\\
Median {\bf A$_{\bf 1.87 \mu m}$} & 1.47 & 1.46 & 1.47 & -- & --\\
Maximum {\bf A$_{\bf 1.87\mu m}$} & 1.71 & 1.71 & 1.69 & 2.27 & 2.32\\
Standard Deviation  & 0.11 & 0.13 & 0.11 & --  & -- \\
Area ($\Box$\arcsec) & 185 & 107 & 120 & -- & -- \\
& & & & & \\
\hline
& & & & & \\
\footnotemark[1]Median {\bf A$_{\bf V}$} & 44.6 & 44.5 & 44.7 & -- & --\\
\footnotemark[1]Maximum {\bf A$_{\bf V}$} &51.9 & 52.1 & 51.5 & 69.1 & 70.5\\
Standard Deviation & 3.2 & 3.9 & 3.4 &  -- & --\\
& & & & &\\
\hline
& & & & &\\
\footnotemark[2]Median {\bf A$_{8\mu m}$} & 1.1 & 1.1 & 1.1 & 1.82\footnotemark[3] & 1.85\footnotemark[3]\\
\footnotemark[4]Serabyn et al. {\bf A$_{12.8\mu m}$} & 1.0 & 1.1 & 1.3 & 3.3 & 3.3\\
& & & & & \\
\hline

\footnotetext[1]{ Calculated using the extinction law of \cite{Nishiyama08}}
\footnotetext[2]{ Calculated using the extinction law of \cite{Nishiyama09}}
\footnotetext[3]{ For region D, the maximum extinction was calculated instead of the median}
\footnotetext[4]{ From \cite{Serabyn92}}
\end{tabular}
\label{extinction}
\end{table}

\end{document}